\documentclass[11pt]{article}
\usepackage{bm}
\usepackage{latexsym}
\usepackage{amsfonts,amsbsy}
\def\Gammabol{{\stackrel{\circ}{\Gamma}}{}}

\def\Rbol{{\stackrel{\circ}{R}}}

\def\be{\begin{equation}}
\def\ee{\end{equation}}
\def\ba{\begin{eqnarray}}
\def\ea{\end{eqnarray}}

\begin{document}

\title{Doing without the Equivalence Principle\footnote{Talk presented at the {\it Tenth
Marcel Grossmann Meeting}, July 20 to 26, 2003, Rio de Janeiro, Brazil.}}

\author{R. Aldrovandi, J. G. Pereira and K. H. Vu \\
Instituto de F\'{\i}sica Te\'orica, 
Universidade Estadual Paulista \\
Rua Pamplona 145,
01405-900\, S\~ao Paulo SP, Brazil}

\date{}  

\maketitle

\begin{abstract}
In Einstein's general relativity, geometry replaces the concept of force in the 
description of the gravitation interaction. Such an approach rests on the universality
of free-fall---the weak equivalence principle---and would break down without it. On the other
hand, the teleparallel version of general relativity, a gauge theory for the translation
group, describes the gravitational interaction by a force similar to the Lorentz
force of electromagnetism, a non-universal interaction. It is shown that, similarly to the
Maxwell's description of electromagnetism, the teleparallel gauge approach provides a
consistent theory for gravitation even in the absence of the weak equivalence principle.
\end{abstract}

\section{Introduction}

Geometry replaces force in Einstein's general relativistic description of gravitation.
Such a geometric description of gravitation is possible because of the universality of free
fall, or the {\em weak} equivalence principle, which establishes the equality of {\em
inertial} and {\em gravitational} masses. Only a universal interaction can be described by a
{\em geometrization} of spacetime, by which all particles of nature, independently of their
internal constitution, feel gravitation the same and, for a given set of initial conditions,
follow the same trajectory. 

On the other hand, the teleparallel equivalent of general relativity \cite{Ham02}, or
teleparallel gravity for short \cite{obs}, a gauge theory for the Abelian translation group,
describes the gravitational interaction, not through a geometrization of spacetime, but by a
force similar to the Lorentz force of electromagnetism, a non-universal interaction. The
question then arises whether, similarly to the Maxwell's description of electromagnetism, the
teleparallel gauge approach is able to give a consistent theory for gravitation even in the
absence of the weak equivalence principle \cite{wep}. The basic purpose of these notes is to
provide an answer to this question. We begin with a review of the fundamentals of
teleparallel gravity.

\section{Teleparallel Gravity: Fundamentals}

Teleparallel gravity is a gauge theory \cite{PR} for the group of translations \cite{sp1}.
These translations take place on the Minkowski tangent space (fiber) to each point of
spacetime (base space).\footnote{We use the Greek alphabet $\mu, \nu, \rho,
\dots = 0, 1, 2, 3$ to denote spacetime indices, and the Latin alphabet $a, b, c, \dots = 0,
1, 2, 3$ to denote anholonomic indices related to the tangent Minkowski spaces, whose metric
is chosen to be $\eta_{a b} = {\rm diag} (+1, -1, -1, -1)$.} The gauge potential of
teleparallel gravity is a 1-form with values in the Lie algebra of the translation group, 
\be
B_\mu = B^a{}_\mu \, P_a,
\ee
with $P_a = \partial_a$ the generators of infinitesimal translations. The corresponding field
strength is
\be
F^a{}_{\mu \nu} = \partial_\mu B^a{}_\nu - \partial_\nu B^a{}_\mu.
\label{tfs}
\ee
A gauge transformation is defined as a local translation of the tangent space coordinates,
\be
x^{\prime a} = x^a + \alpha^a,
\label{transx}
\ee
with $\alpha = \alpha(x^\mu)$ the parameter transformation. Under such a transformation,
the gauge potential $B^a{}_\mu$ behaves as
\be
B^{\prime a}{}_\mu = B^a{}_\mu - \partial_\mu \alpha^a,
\label{transB}
\ee
which leaves the field strength invariant:
\be
F^{\prime a}{}_{\mu \nu} = F^a{}_{\mu \nu}.
\ee
The gauge potential appears in teleparallel gravity as the nontrivial part of the tetrad
field \cite{sp1}:
\be
h^a{}_\mu = \partial_\mu x^a + B^a{}_\mu.
\label{tetrada}
\ee
It is then an easy task to verify that the tetrad is also invariant:
\be
h^{\prime a}{}_\mu = h^a{}_\mu.
\ee
It is important to mention that tangent space indices are raised and lowered with $\eta_{a
b}$, and the spacetime indices are raised and lowered with the spacetime metric
\be
g_{\mu \nu} = \eta_{a b} \; h^a{}_\mu \; h^b{}_\nu.
\label{gmn}
\ee
It is also important to remark that, although depending on the translational gauge potential
$B^a{}_\mu$, the metric tensor does not play any dynamical role in teleparallel gravity.

The tetrad (\ref{tetrada}) gives rise to the Weit\-zen\-b\"ock metric-preserving connection
\begin{equation}
\Gamma^{\rho}{}_{\mu\nu} = h_{a}{}^{\rho} \partial_{\nu} h^{a}{}_{\mu},
\label{carco}
\end{equation}
which has vanishing curvature, and whose non-vanishing torsion
\begin{equation}
T^{\rho}{}_{\mu\nu} = \Gamma^{\rho}{}_{\nu\mu} - 
\Gamma^{\rho}{}_{\mu\nu}
\label{tor}
\end{equation}
coincides with the gauge field strength:
\be
T^{\rho}{}_{\mu\nu} = h_{a}{}^{\rho} F^a{}_{\mu \nu}.
\label{thf}
\ee
That connection can be decomposed as
\begin{equation}
\Gamma^{\rho}{}_{\mu\nu} = \Gammabol^{\rho}{}_{\mu\nu} 
+ K^{\rho}{}_{\mu\nu},
\label{rela}
\end{equation}
where $\Gammabol^{\rho}{}_{\mu\nu}$ is the Christoffel connection of the metric
$g_{\mu\nu}$, and
\begin{equation}
K^{\rho}{}_{\mu \nu} = \frac{1}{2} \left( 
T_{\mu}{}^{\rho}{}_{\nu} + T_{\nu}{}^{\rho}{}_{\mu} 
- T^{\rho}{}_{\mu \nu} \right)
\label{contorsion}
\end{equation}
is the contortion tensor. We remark that curvature and torsion are properties of a
connection, not of the space on which it is defined \cite{livro}. Notice, for example, that
the Christoffel and the Weitzenb\"ock connections, which are connections presenting different
curvature and torsion tensors, are defined on the very same spacetime metric manifold.

The Lagrangian of the teleparallel equivalent of general relativity is \cite{maluf}
\begin{equation}
{\mathcal L}_G = \frac{c^{4} h}{16\pi G} \, S^{\rho\mu\nu}\,T_{\rho\mu\nu} +
{\mathcal L}_M,
\label{gala}
\end{equation}
where $h = {\rm det}(h^{a}{}_{\mu})$, ${\mathcal L}_M$ is the Lagrangian of a source
field and
\begin{equation}
S^{\rho\mu\nu} = - S^{\rho\nu\mu} = \frac{1}{2} 
\left[ K^{\mu\nu\rho} - g^{\rho\nu}\,T^{\sigma\mu}{}_{\sigma} 
+ g^{\rho\mu}\,T^{\sigma\nu}{}_{\sigma} \right]
\label{S}
\end{equation}
is a tensor written purely in terms of the Weitzenb\"ock connection. Variation with respect
to the gauge potential $B^a{}_\mu$ leads to the field equation \cite{sp2}
\begin{equation}
\partial_\sigma(h S_\lambda{}^{\rho \sigma}) -
\frac{4 \pi G}{c^4} \, (h t_\lambda{}^\rho) =
\frac{4 \pi G}{c^4} \, (h {\mathcal T}_\lambda{}^\rho),
\label{eqs1}
\end{equation}
where
\begin{equation}
h \, t_\lambda{}^\rho = \frac{c^4 h}{4 \pi G} \, S_{\mu}{}^{\rho \nu}
\,\Gamma^\mu{}_{\nu\lambda} - \delta_\lambda{}^\rho \, {\mathcal L}_G
\label{emt}
\end{equation}
is the energy-momentum pseudotensor of the gravitational field, and ${\mathcal
T}_\lambda{}^\rho = {\mathcal T}_a{}^\rho \, h^a{}_\lambda$ is the symmetric \cite{weinberg}
energy-momentum tensor of the source field, with
\be
h \, {\mathcal T}_a{}^\rho = -\,
\frac{\delta {\mathcal L}_M}{\delta B^a{}_\rho} \equiv -\,
\frac{\delta {\mathcal L}_M}{\delta h^a{}_\rho}.
\ee
A solution of the gravitational field equation (\ref{eqs1}) is an explicit form of the
gravitational gauge potential $B^a{}_\mu$.

When the weak equivalence principle is assumed to be true, teleparallel gravity turns
out to be equivalent to general relativity. In fact, up to a divergence, the Lagrangian
(\ref{gala}) is equivalent to the Einstein-Hilbert Lagrangian 
\be
{\mathcal L}_G = \frac{c^{4} h}{16\pi G} \; \Rbol,
\ee
with $\Rbol$ the scalar curvature of the Christoffel connection. Accordingly, the
teleparallel field equation (\ref{eqs1}) is found to coincide with Einstein's equation
\be
\Rbol_\lambda{}^\rho - \frac{1}{2} \delta_\lambda{}^\rho \Rbol =
\frac{8 \pi G}{c^4} \, {\mathcal T}_\lambda{}^\rho 
\label{eeq}
\ee
where $\Rbol_\lambda{}^\rho$ is the Ricci curvature of the Christoffel connection. In what
follows, we are going to see what happens when the weak equivalence principle is supposed not
to hold, that is, when the gravitational mass $m_g$ and the inertial mass $m_i$ are assumed
not to coincide. It is important to make it clear that, although there are many controversies
related to the equivalence principle \cite{notabenne}, it is not our intention here to
question its validity, but simply verify whether the teleparallel description of gravitation
requires or not its existence.

\section{Teleparallel Force Equation}

Analogously to electromagnetism \cite{landau}, the action integral of a spinless particle in a
gravitational field $B^{a}{}_{\mu}$ is given by
\be
S = \int_{a}^{b} \left[ - m_i \, c \, d\sigma -
m_g \, c \, B^{a}{}_{\mu} \, u_{a} \, dx^{\mu} \right],
\label{acao1}
\ee
where $d\sigma = (\eta_{a b} dx^a dx^b)^{1/2}$ is the Minkowski tangent-space invariant
interval, and $u^a$ is the particle four-velocity seen from the tetrad frame,
necessarily anholonomic when expressed in terms of the {\it spacetime} line element
$ds$. It should be noticed however that, in terms of the {\it tangent-space} line
element $d \sigma$, it is holonomic \cite{wep}:
\be
u^a = \frac{d x^a}{d \sigma}.
\label{native}
\ee
The first term in action (\ref{acao1}) represents a free particle with inertial mass $m_i$.
The second, its coupling to the gravitational field through its gravitational mass $ m_g$.
We remark that such a decomposition of the action turns up in gauge theories, but  not in
general relativity.

Variation of the action (\ref{acao1}) yields
\ba
\delta S = \int_{a}^{b} m_i c \Big[ \Big( \partial_\mu x^a +
\frac{m_g}{m_i} \; B^a{}_\mu \Big) \frac{d u_a}{d s}
&-& \frac{m_g}{m_i} \; (\partial_\mu B^a{}_\rho -
\partial_\rho B^a{}_\mu ) u_a \, u^\rho \nonumber  \Big]
\delta x^\mu \, ds,
\label{delta1}
\ea
where
\be
u^\mu = \frac{d x^\mu}{ds} \equiv h^\mu{}_a \, u^a
\label{ust}
\ee
is the particle four-velocity, with $ds$ = $(g_{\mu \nu} dx^\mu dx^\nu)^{1/2}$ the
spacetime invariant interval. Using the definition (\ref{tfs}) for the field
strength, we are then left with
\be
\delta S = \int_{a}^{b} m_i c \Big[ \Big( \partial_\mu x^a +
\frac{m_g}{m_i} \; B^a{}_\mu \Big) \frac{d u_a}{d s}
- \frac{m_g}{m_i} \; F^a{}_{\mu \rho} \; u_a \, u^\rho \Big]
\delta x^\mu \, ds.
\label{delta3}
\ee
From the invariance of the action and the arbitrariness of $\delta x^\mu$, it then follows
the force equation
\be
\left( \partial_\mu x^a +
\frac{m_g}{m_i} \; B^a{}_\mu \right) \frac{d u_a}{d s} =
\frac{m_g}{m_i} \; F^a{}_{\mu \rho} \; u_a \, u^\rho.
\label{eqmot2}
\ee
We see clearly from this equation that the teleparallel field strength $F^a{}_{\mu \rho}$
plays the role of gravitational force. Similarly to the electromagnetic Lorentz force
equation, which depends on the relation $e/m_i$, the gravitational force equation depends
explicitly on the relation ${m_g}/{m_i}$ of the particle. Notice furthermore that, due to the
presence of $m_g/m_i$ multiplying the gauge potential, the term between parentheses in the
left-hand side of the above equation of motion is {\em not} the tetrad field.

The crucial point now is to observe that, although the equation of motion (\ref{eqmot2})
depends explicitly on the ratio $m_g/m_i$ of the particle, neither $B^a{}_\mu$ nor
$F^a{}_{\rho \mu}$ depends on this relation. This means essentially that the teleparallel
field equation (\ref{eqs1}) can be consistently solved for the gravitational potential
$B^a{}_\mu$, which can then be used to write down the equation of motion (\ref{eqmot2}),
independently of the validity or not of the weak equivalence principle. The basic conclusion
is that teleparallel gravity is able to describe the motion of a particle with $m_g \neq m_i$.
Accordingly, the gauge potential $B^a{}_\mu$ can be considered as the fundamental field
representing gravitation. 

\section{Relation with Geodesics}

According to teleparallel gravity, even when $m_g \neq m_i$, the tetrad is still given by
(\ref{tetrada}), and the spacetime indices are raised and lowered with the metric
(\ref{gmn}). Then, by using relations (\ref{thf}) and (\ref{rela}), as well as the identity
\be
T^\lambda{}_{\mu \rho} \, u_\lambda \, u^\rho = - K^\lambda{}_{\mu \rho}
\, u_\lambda \, u^\rho,
\ee
the force equation (\ref{eqmot2}) can be rewritten in the form
\be
\frac{d u_\mu}{ds} - \Gammabol^\lambda{}_{\mu \rho} \, u_\lambda \, u^\rho =
\left(\frac{m_g - m_i}{m_g} \right) \partial_\mu x^a \, \frac{d u_a}{d s} .
\label{eqmot6}
\ee
Notice that a violation of the weak equivalence principle produces a deviation from the
geodesic motion, proportional to the difference between the gravitational and
inertial masses. Notice also that, due to the assumed non-universality of free
fall, there is no local coordinate system in which the gravitational effects are absent.

Now, a geometric description for the gravitational interaction of a particle with  $m_g \neq
m_i$ can be obtained by assuming the new tetrad
\be
\bar{h}^a{}_\mu = \partial_\mu x^a + \frac{m_g}{m_i} \; B^a{}_\mu,
\label{tetrada2}
\ee
defining a new spacetime metric tensor $\bar{g}_{\mu \nu} = \eta_{a b} \; \bar{h}^a{}_\mu \;
\bar{h}^b{}_\nu$  and interval  $d\bar{s}^2 = \bar{g}_{\mu \nu} \, dx^\mu dx^\nu$. Notice
that this tetrad is not gauge invariant, as can be seen from Eqs.~(\ref{transx}) and
(\ref{transB}). Furthermore, the relation between the gravitational field strength and
torsion turns out to be in this case
\be
\frac{m_g}{m_i} \; F^a{}_{\mu \rho} = \bar{h}^a{}_\lambda \, \bar{T}^\lambda{}_{\mu \rho}.
\label{fstor2}
\ee
It is then an easy task to verify that, for a fixed relation $m_g/m_i$, the equation of motion
(\ref{eqmot2}) is equivalent now to the geodesic equation
\be
\frac{d \bar{u}_\mu}{d\bar{s}} - {\bar{\Gamma}}{}^\lambda{}_{\mu \rho}
\, \bar{u}_\lambda \, \bar{u}^\rho = 0,
\label{eqmot7}
\ee
where $\bar{u}_\mu \equiv d x_\mu/d \bar{s} = \bar{h}^a{}_\mu u_a$, and
$\bar{\Gamma}{}^{\rho}{}_{\mu\nu}$ is the Christoffel connection of the metric
$\bar{g}_{\mu \nu}$. This equation can also be obtained from the action integral
\be
\bar{S} = -\, m_i \ c \int_a^b d\bar{s},
\ee
which has the usual general relativity form. However, the solution of the corresponding
Einstein's field equation
\be
\bar{R}_{\mu \nu} - \frac{1}{2} \, \bar{g}_{\mu \nu} \bar{R} =
\frac{8 \pi G}{c^4} \, \bar{\mathcal T}_{\mu \nu}
\label{e2}
\ee
would  depend on the ratio $m_g/m_i$ of the test particle. This means essentially that the
resulting gravitational theory is inconsistent in the sense that test particles with different
ratios $m_g/m_i$ would require connections with different curvatures to keep all equations
of motion given by geodesics, as required by a geometric description.

\section{Final Remarks}

In Einstein's general relativity, a theory fundamentally based on the universality of
free fall (or on the weak equivalence principle), geometry replaces the concept of
gravitational force. As a consequence, all equations of motion are necessarily given by
geodesics. This theory has been confirmed by all experimental tests at the classical level,
but any violation of the principle would lead to its ruin. On the other hand, the teleparallel
equivalent of general relativity does not geometrize the interaction, but shows gravitation
as a gauge force quite analogous to the Lorentz force of electrodynamics. It is, therefore,
not committed with geodesics. As a consequence of this fundamental difference, it is able to
describe the gravitational interaction in the absence of universality just as Maxwell's
gauge theory is able to describe the non-universal electromagnetic interaction \cite{wep}. In
spite of the equivalence with general relativity when the weak equivalence principle is
assumed to be true, it can  be considered as a more fundamental theory as it dispenses with
one assumption. Notice in this connection that the equivalence principle is frequently said
to preclude the definition of a local energy-momentum density for the gravitational
field \cite{gravitation}. Although this is a true assertion in the context of general
relativity, it has already been shown that a tensorial expression for the gravitational
energy-momentum density is possible in the context of teleparallel gravity \cite{sp2}.

In the teleparallel approach, the fundamental field describing gravitation is the
translational gauge potential $B^a{}_\mu$ \cite{global}. Quantization of the gravitational
field, therefore, should be carried out on $B$ and not on the tetrad or on the metric. In
addition, gravitational waves should be interpreted as $B$ waves and not as metric waves. An
aspect which could lead to an eventual test by laboratory experiment is a deep difference
between spin-2 and spin-1 mediating fields: an interaction mediated by a vector field would
give opposite signs for matter-matter and matter-antimatter interactions, while a mediating
spin-2 field gives the same sign for both \cite{kibble}. Another important consequence refers
to a fundamental problem of quantum gravity, namely, the conceptual difficulty of reconciling
{\it local} general relativity with {\it non-local} quantum mechanics, or equivalently, of
reconciling the local character of the equivalence principle with the non-local character of
the uncertainty principle \cite{exp}. As teleparallel gravity can be formulated independently
of the equivalence principle, the quantization of the gravitational field may possibly
appear more consistent if considered in the teleparallel picture. Finally, we would like to
remark that, on the strength of our results, provided the teleparallel approach to
gravitation be used, the old suggestion made by Synge about the equivalence principle,
namely, that the {\it midwife be now buried with appropriate honours} \cite{synge}, can thus
be realized.

\section*{Acknowledgments}
The authors would like to thank FAPESP-Brazil, CNPq-Brazil, and CAPES-Brazil for
financial support.

\end{document}